\documentclass[%
 aip,
rsi,%
 amsmath,amssymb,
 reprint,%
]{revtex4-1}

\usepackage{graphicx}
\usepackage{dcolumn}
\usepackage{bm}
\usepackage{wasysym}

\usepackage{color,soul}
\newcommand{\specialcell}[2][c]{%
  \begin{tabular}[#1]{@{}c@{}}#2\end{tabular}}

\begin{document}

\preprint{AIP/123-QED}

\title{A geometry-independent moment correction method for the Magnetic Property Measurement 3 Superconducting Quantum Interference Device-Vibrating Sample Magnetometer}

\author{C. O. Amorim}
 \affiliation{Physics Department and  CICECO, University of Aveiro, 3810-193 Aveiro, Portugal}
 \author{F. Mohseni}
 \affiliation{Physics Department and  CICECO, University of Aveiro, 3810-193 Aveiro, Portugal}
 \author{V. S. Amaral}
 \affiliation{Physics Department and  CICECO, University of Aveiro, 3810-193 Aveiro, Portugal}
 \author{J. S. Amaral}
 \affiliation{Physics Department and  CICECO, University of Aveiro, 3810-193 Aveiro, Portugal}

\date{\today}

\begin{abstract}
The sensitivity and automation capabilities of modern superconducting quantum interference device (SQUID) magnetometers are currently unmatched. The measured moment values are, however, prone to deviations from their actual value due to geometric effects, namely sample size, shape, and radial offset. This is well known, and a knowledgeable operator will correct measured moment values taking these effects into account. 

The current procedure for the Magnetic Property Measurement 3 (MPMS3) magnetometer is based on an available simulation tool, valid for both Vibrating Sample Magnetometer (VSM) and Direct Current (DC) methods. Still, determining the correction factor requires samples with well-defined geometric shapes together with accurate sample dimensions and the usually difficult to determine radial offset. Additionally, at the moment, there is not a proper solution to correct geometry effects of irregular shaped samples. 

In this work, we find a systematic relation between the difference between the VSM and DC measurements and their corresponding correction factors for MPMS3 SQUID-VSM device. This relation follows a clear trend, independent of sample size, shape or radial offset, for a given pair of DC scan length and VSM amplitude values. Exploiting this trend, a geometry-independent correction method is here presented and validated by measurements of metallic Fe powder using a far from optimal sample mounting. 
%
\end{abstract}

\pacs{Valid PACS appear here}
\maketitle

\section{Introduction}
The usage of superconducting quantum interference devices is an increasingly popular approach to magnetometry, mainly to its intrinsic unmatched sensitivity, and to due to the widely available commercial automated devices
\cite{Buchner2018}. These devices are mostly based on second-order gradiometers, and the measured magnetic moments are susceptible to inaccuracies due to geometric and offset effects, mainly for samples which occupy more than 5\% of the gradiometer volume
\cite{Coey2006,new_ref1,new_ref2}. 

For the particular case of the popular Quantum Design (QD) MPMS3 these geometric effects can depend heavily on the sample size, shape and radial offset, thus requiring an adequate correction from the SQUID's operator
\cite{QD_Manual, QD_radial,QD_tabela_erros}. 

Currently the correction method provided by the MPMS3 supplier, consists in a simulation software which provides the adequate correction factors for both the VSM and DC measurement methods, given the sample geometry (solely for cylinders, rectangular prisms, or thin films), as well as its accurate dimensions and radial offset
\cite{QD_1500-020_2014}. 

However, for everyday measurements, the samples may have quite irregular shapes, and the determination of the radial offset is quite difficult, even for regular shaped samples, being based on a best guess, as reported in the QD MPMS3 Application Note 1500-020
\cite{QD_1500-020_2014}. These difficulties can severely limit the accuracy of the measurements, as much as tens of percent for certain given geometries and a bad enough offset determination, even though the high sensitivity of this SQUID magnetometer (as high as 10$^{-8}$ emu). Since the MPMS3 is currently the sole SQUID magnetometer currently sold by Quantum Design, the largest commercial SQUID magnetometer manufacturer, solving these limitations is of great importance to the scientific community.

To overcome these practical limitations, a geometry independent method to correct sample geometries effects on the measured moment of the QD MPMS3 magnetometer is here presented. This method is based on the correlation between the difference between VSM and DC measurements and their respective correction factors, obtained from the QD MPMS3 Sample Geometry Simulator (SGS)
\cite{QD_1500-020_2014}.        

\section{The Correction Method}
In this manuscript we propose a method based on the assumption that, for each pair of DC scan lengths and VSM amplitudes, there is a unique function of correction factors, $\alpha$, which depends only on the relative difference between the VSM and DC measurements, defined by the ratio $x = \left( \text{M}_{\text{VSM}} - \text{M}_{\text{DC}} \right) / \text{M}_{\text{VSM}}$.

To confirm this conjecture, several hundred of points were simulated using the QD MPMS3 SGS
\cite{QD_1500-020_2014}. For each pair of DC scan lengths and VSM amplitudes, several shapes, sizes and radial offsets were simulated, using the available geometries: cylinders, rectangular prisms, and parallel and perpendicular thin films. The simulations scanned all the available parameters, namely: heigh, diameter, width, depth and radial offset, within values which respected the physical limits of the sample chamber ($\diameter \approx 8$\,mm). 

\begin{figure}
\includegraphics[width=\columnwidth]{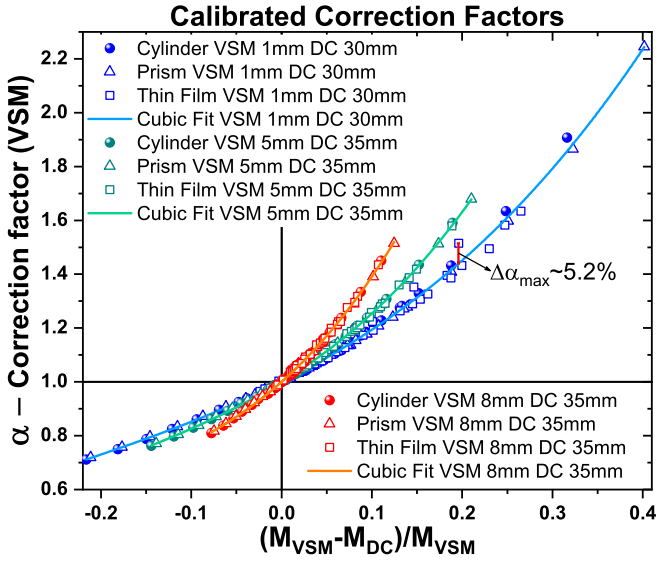}
\caption{\label{Fig1} Simulated correction factors as a function of the relative difference $x = \left( \text{M}_{\text{VSM}} - \text{M}_{\text{DC}} \right) / \text{M}_{\text{VSM}}$ for three different pairs of DC scan lengths and VSM amplitudes.}
\end{figure}

After doing the simulation of the correction factors for several geometries, it is possible to confirm that indeed, each pair of DC scan lengths and VSM amplitudes follows a unique trend, independently of the considered geometry and radial offset, as it is shown in figure \ref{Fig1}. Figure \ref{Fig1} also shows that this function, $\alpha (x)$, can be empirically fitted with great success to the cubic polynomial presented in equation \ref{Eq1}. 

\begin{equation}\label{Eq1}
\alpha(x) = 1 + Ax + Bx^2 + Cx^3 ,
\end{equation}

Table \ref{table1} presents the fit parameters from this empirically fitted function, for the pairs of DC scan lengths and VSM amplitudes presented in figure \ref{Fig1}. An expanded table, with the fit functions of additional pairs of DC scan lengths and VSM amplitudes, is also supplied by our group, whose database will be continually increased 
\cite{LINK}. Figure \ref{Fig1} shows that, for the considered simulated parameters domain, there is a decrease of the uncertainty for higher VSM amplitudes, as reported by QD MPMS3 Application Note 1500-020 
\cite{QD_tabela_erros}.     

\begin{table}
\caption{\label{table1}Fitted $A$, $B$ and $C$ parameters  of the VSM correction factor function, $\alpha(x)$, for the pairs of VSM amplitudes and DC scan lengths presented at figure \ref{Fig1}.}
\begin{ruledtabular}
\begin{tabular}{ccccc}
DC  & VSM  &  & {\large $\alpha(x)$} & \\
scan length & amplitude & $A$ & $B$ & $C$\\
\hline
30 mm & 1 mm & $1.68 \pm 0.02$ & $2.27 \pm 0.09$ & $3.1 \pm 0.3$\\
35 mm & 5 mm & $2.068 \pm 0.007$  & $4.00 \pm 0.05$ & $7.1 \pm 0.4$\\
35 mm & 8 mm & $2.98 \pm 0.02$ & $8.3 \pm 0.2$ & $10 \pm 3$\\
\end{tabular}
\end{ruledtabular}
\end{table}

The determination of each correction factor function, $\alpha(x)$, allows the user to make the required geometric corrections without the need of actually measuring the dimensions of the sample, or estimating its proper radial offset
\cite{Morrison}. 

To perform this correction, the user only needs to make additional pairs of measurements using the VSM amplitude and DC scan length compatible with its measurement sequence, in order to obtain the value of the variable $x$. Knowing the measured $x$ value, let's call it $X_1$, the user can then chose the $\alpha(x)$ function corresponding to the pair of DC scan lengths and VSM amplitudes which he used, and correct all the following measurements using equation \ref{Eq2}: 

\begin{equation}\label{Eq2}
M_\text{real} = \frac{M_\text{measured}}{\alpha(X_1)}
\end{equation}
where $M_\text{measured}$ is the measured magnetization, and $M_\text{real}$ is the actual magnetization of the sample.

Note that a similar approach can be used to correct DC measurements, replacing the $y$ axis presented in figure \ref{Fig1} by the correction factors of the DC scan. Several of these correction fits can also be found in the same database of the VSM corrections
\cite{LINK}.

\section{Application to metallic Iron powder data}
To validate this sample geometry correction method, several gelatin capsules containing different volumes of Fe powder (99\% purity from Sigma Aldrich) were measured with a QD MPMS3 SQUID-VSM at 300K, using magnetic fields ranging from 0--30\,kOe. The powder volumes were chosen to be incrementally smaller, from 2.745\,mm$^3$ (21.614\,mg) to 0.237\,mm$^3$ (1.869\,mg), making its geometric distribution at the bottom of the capsule completely irregular and difficult to determine. The magnetization measurements were performed using both the VSM and DC methods, with an amplitude of 1\,mm and a scan length of 30\,mm respectively.

\begin{figure}[h]
\includegraphics[width=\columnwidth]{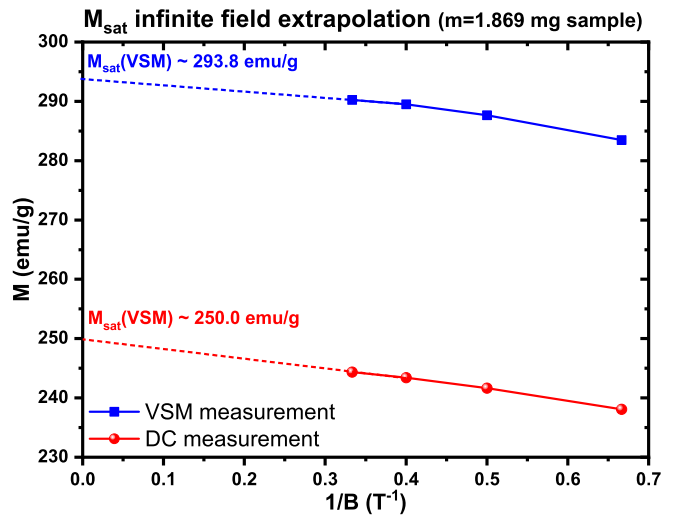}
\caption{\label{Fig2} Illustrative figure of the determination of the magnetization saturation through the extrapolation of the $M(1/B)$ plot to infinite magnetic fields. This figure corresponds to the Fe sample with $m = 1.869$\,mg.}
\end{figure}

The VSM and DC saturation magnetizations are presented in table \ref{tab:table3}, and result from the extrapolation of the $M(1/B)$ curves (figure \ref{Fig2}). Table \ref{tab:table3} shows that for the  VSM measurements there are significant deviations in the saturation magnetization of Fe, when compared with the expected Fe saturation magnetization at 298.15\,K, $\text{M}_{\text{sat,T=298.15K}} = 218.1$\,emu/g
\cite{Fe_saturation2}, as high as 34.7\%.

\begin{table*}
\caption{\label{tab:table3} Measured saturation magnetization and respective corrected values of several Fe powder samples.}
\begin{ruledtabular}
\begin{tabular}{cccccccc}
\specialcell{mass\\($\pm$ 0.001\,mg)} & \specialcell{$\text{M}_{\text{sat}}$ (VSM)\\(emu/g)} & \specialcell{Experimental\\deviation} &\specialcell{$\text{M}_{\text{sat}}$ (DC)\\(emu/g)} & $\frac{\text{M}_{\text{VSM}} - \text{M}_{\text{DC}}}{\text{M}_{\text{VSM}}}$ & \specialcell{$\alpha$} & \specialcell{Corrected\\$\text{M}_{\text{sat}}$ (emu/g)} & \specialcell{Corrected\\deviation}\\
\hline
 1.869  & $293.8$  & $+34.7\%$ & $250.0$ & 14.9\%  & $1.31$   & $224.1$ & $+2.8\%$ \\
 3.625  & $255.1$  & $+17.0\%$ & $229.0$ & 10.2\%  & $1.199$ & $212.8$ & $-2.4\%$ \\
 9.793  & $286.4$  & $+31.3\%$ & $242.0$ & 15.5\%  & $1.32$   & $215.9$ & $-1.0\%$ \\
21.614  & $271.4$  & $+24.4\%$ & $235.8$ & 13.1\%  & $1.266$ & $214.3$ & $-1.7\%$ \\
\end{tabular}
\end{ruledtabular}
\end{table*}

On the other hand, using the proposed geometry method, we are able to consistently decrease the experimental magnetization deviations from the expected Fe saturation magnetization to values $\leq 2.8\%$. This deviation from the expected value is lower than the maximum $5.2\%$ deviation of the simulated data from its corresponding cubic fit. Nevertheless, this $2.8\%$ deviation is still almost $13\times$ better than the initial measured saturation magnetization. 

\subsection*{Application to a general sample}
The presented correction method solely corrects geometry and offset effects, hence several concerns must be addressed to reduce other uncertainty variables such as centering issues or sample holder background and heterogeneity effects. Therefore, we suggest a few steps to implement our correction method successfully.

First the user should be careful during the sample mounting, using the cleanest possible conditions, to avoid external contaminations. If the sample has more than one piece (or if it is a powder), the user should warrant that the pieces are placed together in order to minimize a heterogeneous distribution in the sample holder/capsule. Additionally, the amount of sample should be chosen adequately to minimize the effects of the sample holder background.

For VSM mode measurements, the user should choose the VSM amplitude to be as high as possible for the expected range of measured magnetization. This means that preliminary measurements might be required to gage the order of magnitude of the sample magnetization. A high VSM amplitude will have a lower spread of corrections values, as shown in figure \ref{Fig1}, allowing a more accurate geometry and offset correction. 

Before performing the magnetization measurements, the user should perform at least one pair of VSM and DC measurements using the same VSM amplitude and/or DC scan length that will be used in the magnetization curve measurements. This pair of VSM and DC measurements will be used to calculate the ratio $X_1 = \left( \text{M}_{\text{VSM}} - \text{M}_{\text{DC}} \right) / \text{M}_{\text{VSM}}$ required to determine the proper $\alpha(X_1)$, by substitution in equation \ref{Eq1}. The $A$, $B$ and $C$ parameters can be found in Ref. \cite{LINK} for the available pairs of VSM amplitudes and DC scan lengths or can be obtained using the SGS to determine other $\alpha (x)$. 

Finally, the user should correct the measured magnetization curves using equation \ref{Eq2}.

\section{Conclusions}
In this work, we have presented a geometry-independent sample moment correction method which employs both DC and VSM data from the MPMS3 SQUID-VSM magnetometer. Independent of sample shape, dimensions and radial offset, and for a given pair of DC scan length and VSM amplitude values, a single function correlates the moment correction factor to the difference between DC and VSM measurements. 

We show this approach for three pairs of DC/VSM amplitudes with sample geometry parameters that lead to correction factors up to 2.245 for the VSM case (up to 1.343 for the DC case), using hundreds of simulated points. Within the measured dataset, the maximum observed deviation of the correction factor to the cubic fit was $5.5\%$. This deviation is still about $10 \times$ lower than its initial offset. We should stress out that the accuracy of this method are directly limited by the SGS, and that this does not solve other measurement problems, such as the sample holder background or the issues related to highly heterogeneous samples.  

To validate this correction method, we have applied it to the experimental data of Fe powders, in irregular shapes, large radial offset, and a sample volume as low as 0.2374 mm$^{3}$ ($\backsim 1/100$  the Pd calibration sample), to emulate a far from optimal sample mounting scenario. The initial maximum observed deviation to the expected result was of 34.7\%, which was reduced to a maximum of 2.8\% using the reported method. 

While this approach requires both DC and VSM data, its ease of use on irregular shaped samples and unknown radial offset values, can be of an immediate wide application by the magnetism/magnetometry community. Ultimately, a similar comparative measurement approach could be implemented in future second-order gradiometer-based magnetometers as a means to decrease geometry and offset-based inaccuracies. 

\section*{Data Availability Statement}
The data that support the findings of this study are available from the corresponding author upon reasonable request.

\begin{acknowledgments}
This work was developed within the scope of the project CICECO-Aveiro Institute of Materials, UIDB/50011/2020 and UIDP/50011/2020, financed by national funds through the FCT/MEC. JSA acknowledges FCT IF/01089/2015 grant.
\end{acknowledgments}

\bibliography{Geometry}

\end{document}